\documentclass[11pt,a4paper]{article}

\usepackage{bm,bbm,amsfonts,mathrsfs}
\usepackage{graphicx}
\usepackage{hyperref}

\newcommand{\Z}{\mathbb{Z}}

\newcommand{\U}{\mathrm{U}}
\newcommand{\SU}{\mathrm{SU}}

\newcommand{\Tr}{\mathrm{Tr}\,}
\newcommand{\id}{\mathbbm{1}}

\makeatletter

\@addtoreset{equation}{section}
\makeatother

\date{\today}

\begin{document}

\begin{titlepage}

\renewcommand{\thefootnote}{\fnsymbol{footnote}}

\begin{flushright}
{\tt
RIKEN-MP-58
}
\end{flushright}

\vskip10em

\begin{center}
 {\Large {\bf 
 Hofstadter problem in higher dimensions
 }}

 \vskip3em

 {\sc Taro Kimura}%
 \footnote{
 E-mail address: 
 \href{mailto:taro.kimura@cea.fr}
 {\tt taro.kimura@cea.fr}}

 \vskip2em

{\it 
 Institut de Physique Th\'eorique,
 CEA Saclay, 91191 Gif-sur-Yvette, France
 \\ \vspace{.5em}
 Mathematical Physics Laboratory, RIKEN Nishina Center, Saitama 351-0198,
 Japan 
}

 \vskip3em

\end{center}

 \vskip2em

\begin{abstract}
We investigate some generalizations of the Hofstadter problem to
 higher dimensions with Abelian and non-Abelian gauge field
 configurations.
We numerically show the hierarchical structure in the energy spectra
 with several lattice models.
It is also pointed out the equivalence between the $\pi$-flux state and the
 staggered formalism of Dirac fermion.
\end{abstract}

\end{titlepage}

\tableofcontents

\setcounter{footnote}{0}


\section{Introduction}
\label{sec:Intro}

The fractal structure of the energy spectrum for the two-dimensional
magnetic lattice system, known as Hofstadter's
butterfly~\cite{PhysRevB.14.2239}, is one of the most exotic
consequences of the quantum property of the low-dimensionality.
Such a fractal nature of the magnetic system can be observed in a simple
tight-binding model in the presence of the magnetic field.
This Hofstadter problem and its variants have been extensively discussed in
various areas of physics and also mathematics.
More recently it can be realized even in experimental
situations~\cite{Aidelsburger:2013PRL,Miyake:2013PRL}.

In this paper we extend the Hofstadter problem, which is originally
considered in two dimensions, to higher dimensions with not only Abelian,
but also non-Abelian gauge field configuration.
So far there are some attempts to generalize it to the three-dimensional
magnetic system~\cite{JPSJ.59.4384,PhysRevB.44.6842,PhysRevB.45.13488,PhysRevLett.86.1062},
and also to that in non-Abelian gauge potential~\cite{Goldman:2007EPL78,Goldman:2007EPL80,Goldman:2009PRA,Goldman:2009PRL,Cocks:2012PRL}.
But its generalization to much higher dimensional system has been not
yet studied in the literature.
Actually, when we analyse topological matters in the lower dimensional
system, the four-dimensional point of view can be quite useful: topological
insulators/superconductors in two and three
dimensions~\cite{Hasan:2010xy,Hasan:2010hm} are deeply connected to the
four-dimensional QHE~\cite{Zhang:2001xs} through the dimensional
reduction procedure~\cite{Qi:2008ew,1367-2630-12-6-065010}.
The formalism discussed in this paper can be applied to arbitrary even
dimensional lattice system in the presence of the magnetic field.
Based on this formalism, we prove the $\pi$-flux
state~\cite{PhysRevB.37.3774,PhysRevB.39.11538}, which has a gapless
excitation in general, is essentially equivalent to Dirac fermion in
arbitrary dimensions.
Furthermore, when we apply the non-Abelian gauge field, we can obtain a
hierarchical structure in the energy spectrum of the lattice model even
in higher dimensions.

This paper is organized as follows.
In Sec.~\ref{sec:Abelian} we introduce arbitrary even dimensional
lattice models with Abelian gauge field background configuration.
We discuss the corresponding Schr\"odinger equation to the magnetic
system, and obtain Harper's equation.
We then comment on its connection to the non-commutative torus.
We also provide a general proof for the gapless spectrum of the
$\pi$-flux state, by referring to its equivalence to the naively discretized
lattice Dirac fermion.
In Sec.~\ref{sec:nonAbelian} we then consider the Hofstadter problem for
the non-Abelian gauge field configuration.
In particular the four-dimensional $\SU(2)$ theory is investigated as a
fundamental example of the non-Abelian gauge theory.
We show some numerical results of the model, and discuss the effect of
the inhomogeneity of the background flux.
Section~\ref{sec:summary} is devoted to a summary and discussion.

\section{Abelian gauge field models}\label{sec:Abelian}

First generalization of Hofstadter problem is formulated in arbitrary
even dimensions in the presence of $\U(1)$ gauge field.
Before introducing a lattice model, we now consider the following background
$\U(1)$ field configuration for the continuum theory \cite{Smit:1986fn},
\begin{equation}
 F_{2s-1,2s} = \omega_{s} , \quad
 F_{\mu\nu} = 0 \quad \mbox{for} \quad \mbox{otherwise}, \quad
 \left(s = 1, \cdots, r \right) .
\end{equation}
We then apply the higher dimensional version of Landau gauge to this
configuration,
\begin{eqnarray}
 & A_{2s-1}(x) = - \omega_{s} x_{2s}, \qquad
  A_{2s}(x) = 0 . &
  \label{gauge_pot}
\end{eqnarray}
Here the field strengths $\omega_{s}$ are quantized,
\begin{equation}
 \omega_{s} = \frac{2\pi}{L_{2s-1} L_{2s}} n_{s}, \qquad
 n_{s} \in \mathbb{Z},
\end{equation}
and thus the topological number is given by
\begin{equation}
 Q = \frac{1}{2^{r}\cdot r!}
  \int d^{2r} x \, \epsilon^{\mu_1\cdots\mu_{2r}} \,
  \Tr F_{\mu_1\mu_2} \cdots F_{\mu_{2r-1},\mu_{2r}}
  = \prod_{s=1}^{r} n_{s} .
\end{equation}
This is regarded as the $r$-th Chern number.


To discuss the Hofstadter spectrum, we then realize these configurations
on the lattice.
The gauge potential (\ref{gauge_pot}) is implemented by introducing the
link variable, which can be regarded as the Wilson line,
\begin{eqnarray}
 & U_{2s-1}(x) = e^{-i\omega_{s} x_{2s} }, \qquad
  U_{2s}(x) = 1 . & 
\end{eqnarray}
Precisely speaking, we have to assign appropriate boundary conditions
even for $U_{2s}(x)$ \cite{Smit:1986fn}.
In this case the field strength is restricted to the interval
$(0,2\pi)$ due to the lattice discretization
\cite{Panagiotakopoulos:1984qi,Phillips:1986us}.
Furthermore they are characterized by the following fractions,
\begin{equation}
 \frac{p_{s}}{q_{s}} \equiv 
 \frac{n_{s}}{\tilde L_{2s-1} \tilde L_{2s}}, \qquad
\end{equation}
Here $\tilde L_i$ is related to the system size as $L_i = \tilde L a$
where $a$ is the lattice spacing.
Note that they satisfy $0 < p_i / q_i < 1$.
This fraction plays an essential role in the interesting spectrum of
the model we discuss below.

\subsection{Tight-binding model}\label{sec:tight}

The lattice Hamiltonian with this background configuration is
defined as
\begin{equation}
 \mathcal{H}_{\mathrm{tight}} = \sum_{x,\mu} 
  \left[
    c_{x+\hat\mu}^\dag U_\mu(x) c_{x}
   + c_{x}^\dag U_\mu^\dag(x) c_{x+\hat\mu}
  \right].
  \label{Ham}
\end{equation}
This is just the tight-binding hopping model, which describes the
non-relativistic particle with the background {\em magnetic} field.
Introducing the state $|\psi\rangle = \sum_x \psi_x c_x^\dag |0\rangle$,
we obtain the corresponding Schr\"odinger equation from the Hamiltonian
(\ref{Ham}) written in a second quantized form,
\begin{equation}
 \sum_{\mu=1}^{2r}
  \left[
   U_{\mu}(x-\hat{\mu}) \psi_{x+\hat\mu}
   + U_{\mu}^\dag(x) \psi_{x-\hat\mu}
  \right]
  = E \psi_{x}.
  \label{Schrodinger}
\end{equation}

We then solve the equation (\ref{Schrodinger}) by taking Fourier
transformation.
Remark the translation symmetry of this model is slightly modified from
the usual lattice model due to the background field,
\begin{equation}
 x_{2s-1} \sim x_{2s-1} + 1, \quad
 x_{2s} \sim x_{2s} + q_{s} .
\end{equation}
Thus, writing the coordinate as $x_{2s} = q_s y_{s} + z_{s}$, the wavefunction
is Fourier transformed as
\begin{equation}
 \psi_x = \frac{1}{V} \sum_{k_1, \cdots, k_{2r}} 
  \exp \left[
	i \sum_{s=1}^{r} 
	\left( k_{2s-1} x_{2s-1} + q_s k_{2s} y_{s} \right)
       \right]
  \tilde \psi_{z_1, \cdots, z_{r}}(k_1, \cdots, k_{2r})
\end{equation}
where $V$ stands for the effective volume of the system, $V = L_1 \cdots
L_{2r}/(q_1 \cdots q_{r})$.
The Schr\"odinger equation (\ref{Schrodinger}) is rewritten in this
basis as
\begin{equation}
 \sum_{s=1}^{r} 
  \left[
     \tilde \psi_{z+\hat{s}} 
   + \tilde \psi_{z-\hat{s}}
   + 2 \cos \left( k_{2s-1} - \omega_s z_s \right)
   \tilde \psi_z
  \right]
  = E \tilde \psi_z .
  \label{Harper}
\end{equation}
This is the higher dimensional version of Harper's equation
\cite{0370-1298-68-10-304}.
While an one-dimensional equation is obtained from the magnetic two-dimensional
system, we have a $r$-dimensional equation from the $d=2r$ theory.
Furthermore the matrix size of this higher dimensional Harper's equation
is $N_q \times N_q$ where $N_q \equiv q_1 \cdots q_{r}$.
This means that the number of energy bands is just given by $N_q$, and
there possibly exist $N_q-1$ energy gaps.
To discuss these energy gaps, one has to investigate a transfer matrix
and its spectral curve associated with Harper's equation (\ref{Harper}).
In this case, however, it is impossible to show these energy
bands are totally gapped with the same argument as the two-dimensional
situation:
we cannot apply a naive transfer matrix method, since there are still
$r$ directions in the equation (\ref{Harper}).
Therefore it is difficult to obtain the corresponding {\em Hofstadter's
butterfly} by diagonalizing Harper's equation
(\ref{Harper}): its {\em wings} are almost disappearing.

Let me comment on the relationship between the non-commutative space and
(\ref{Harper}).
Indeed the higher dimensional tight-binding model, discussed in this
section, can be represented in terms of the non-commutative torus.
The coordinates of the two-dimensional non-commutative torus
$T_\theta^2$ are given by
\begin{equation}
 U V = e^{2\pi i \theta} V U ,
\end{equation}
with $e^{2\pi i \theta}$ being the $q$-th root of unity,
$2\pi\theta \equiv \omega = 2\pi/q$.
They can be written in $q \times q$ matrix forms,
\begin{equation}
 U = \left(      
      \begin{array}{ccccc}
       0 & 1 & & & \\
       & 0 & 1 & & \\
       & & \ddots & \ddots & \\
       & & & 0 & 1 \\
       1 & & & & 0 \\
      \end{array}
     \right), \qquad
 V = \left(
      \begin{array}{ccccc}
       1 & & & & \\
        & e^{i\omega} & & & \\
       & & e^{2i\omega} & & \\
       & & & \ddots & \\
       & & & & e^{(q-1)i\omega} \\
      \end{array}
     \right) .
 \label{NC_torus}
\end{equation}
Based on this description, we can consider $d=2r$ dimensional
non-commutative torus:
\begin{equation}
 U_s V_t = e^{2 \pi i \theta_{st}} V_s U_t
  \qquad \mbox{for} \quad
  s, t = 1, \cdots, r .
\end{equation}
In general, $\theta_{st}$ is a real symmetric matrix.
In the case of (\ref{Harper}), it is simply given by a diagonal
matrix, $2\pi\theta_{st}=\omega_s \delta_{st}$.
This means the non-commutativity on the $2r$-dimensional torus is
introduced to each two-dimensional subspace, $T_\theta^{2r} \to
T_{\theta_1}^2 \times \cdots \times T_{\theta_r}^2$.
Then the associated Hamiltonian can be written as
\begin{equation}
 \mathcal{H}_{\rm tight} = \sum_{s=1}^r
  \left[
   U_s + U_s^\dag + V_s + V_s^\dag
  \right] .
\end{equation}
Although, precisely speaking, we have to include a factor corresponding to the
plane wave, we now omit these factors for simplicity.
This representation means that the non-commutative torus operator plays
a role of the translation operator with the external magnetic field even
in the higher dimensional case.

\subsubsection{$\pi$-flux state}

Let us comment on the $\pi$-flux state in higher
dimensions.%
\footnote{The author is grateful to T.~Misumi for pointing out an
essential connection of this argument to that discussed
in~\cite{Grignani:2000mt}. See also \cite{Creutz:2013ofa}.}
For the $\pi$-flux state~\cite{PhysRevB.37.3774,PhysRevB.39.11538} all
the plaquettes have the same value, $P_{\mu\nu} = -1$, independent of
its position $x$, and directions $\mu$, $\nu$ as
\begin{equation}
 P_{\mu\nu}(x) = U_\mu(x) U_\nu(x+\hat\mu) 
  U_\mu^\dag(x+\hat\nu) U_{\nu}^\dag(x) = -1 .
\end{equation}
Such a configuration is realized by the following $\Z_2 \subset\U(1)$
link variables,
\begin{equation}
 U_\mu(x) = \eta_\mu \equiv (-1)^{x_1 + \cdots + x_{\mu-1}} .
\end{equation}
Then the tight-binding Hamiltonian in the second quantized form 
with this gauge configuration yields
\begin{equation}
 \mathcal{H}_{\rm tight} = \sum_x \sum_{\mu=1}^d \eta_\mu 
  c_x^\dag \left( c_{x+\hat{\mu}} + c_{x-\hat{\mu}}\right) .
  \label{pi-flux}
\end{equation}
We now consider the free field theory for simplicity.
This is almost the same as the staggered Dirac operator
\cite{Kogut:1974ag,Susskind:1976jm,Sharatchandra:1981si}
\begin{equation}
 \mathcal{S}_{\rm staggered} = \sum_x \sum_{\mu=1}^d 
  \frac{1}{2} \eta_\mu 
  \bar\chi_x \left( \chi_{x+\hat{\mu}} - \chi_{x-\hat{\mu}} \right) .
  \label{staggered}
\end{equation}
Actually, by applying the transformation $c_x \to i^{x_1+\cdots+x_d}c_x$,
the Hamiltonian (\ref{pi-flux}) is rewritten as
\begin{equation}
 \mathcal{H}_{\rm tight} = \sum_x \sum_{\mu=1}^d i \, \eta_\mu
  c_x^\dag \left( c_{x+\hat{\mu}} - c_{x-\hat{\mu}}\right) .
\end{equation}
This is just the staggered fermion action (\ref{staggered}), up to a
constant factor.
Note that the staggered Dirac operator is anti-Hermitian while the
tight-binding Hamiltonian is Hermitian.

There is no spinor structure in this formulation, but this staggered
fermion is directly obtained from the naive Dirac fermion through the
spin-diagonalization.
See Appendix \ref{sec:spin_diag} for details.
The sign factor $\eta_\mu = (-1)^{x_1+\cdots + x_{\mu-1}}$ is a
remnant of the gamma matrix.
The staggered fermion enjoys an exact $\U(1)$ chiral symmetry, which is
generated by $\epsilon_x = (-1)^{x_1+\cdots+x_d}$.
This symmetry ensures its gapless excitation.
Therefore the $\pi$-flux state is also generically gapless.

\subsection{Dirac fermion models}\label{sec:Dirac}

We then attempt to extend the Hofsdater's problem to the relativistic
system.
Similar approaches have been seen in the context of graphene
\cite{PhysRevB.73.235118,PhysRevB.74.155415,PhysRevB.74.165411,PhysRevB.75.201404}.

The generic form of the Dirac fermion action is written as
\begin{equation}
 \mathcal{S}_{\mathrm{Dirac}} 
  = \sum_{x,\mu}
  \left[
     \bar\psi_{x+\hat\mu} U_\mu(x) P_\mu^+ \psi_x
   - \bar\psi_{x} U^\dag_\mu(x) P_\mu^- \psi_{x+\hat\mu}
  \right] 
  \equiv \sum_{x,y} \bar \psi_x D_{x,y} \psi_y.
  \label{Ham_Dirac}
\end{equation}
Here $P_\mu^{\pm}$ stands for the spinor matrix, corresponding to a type
of the lattice fermion.\footnote{For example, see \cite{Kimura:2011ik}.}
The simplest one is given by
\begin{equation}
 P_\mu^{\pm} = \frac{1}{2} \gamma_\mu .
\end{equation}
This is just the naive lattice discretization of the relativistic
fermion \cite{Wilson:1974sk}.%
\footnote{
As well known this simple lattice discretization scheme has a problem:
there exist extra massless modes at low energy, which are called the
{\em species doublers}.
To obtain a single chiral fermion we have to implement a much
complicated scheme, e.g. domain-wall fermion, overlap fermion and so on
(see, for example, a recent review \cite{Creutz:2011hy}).
However this problem does not concern our study because it does not
affect on the spectrum of the Dirac fermion.
}
We have other choices for the spinor matrix:
\begin{itemize}
 \item Wilson fermion \cite{Wilson:1974sk}
       \begin{equation}
	P_\mu^{\pm} = \frac{1}{2} ( \gamma_\mu \pm r \id)
       \end{equation}
 \item Karsten--Wilczek fermion \cite{Karsten:1981gd,Wilczek:1987kw}
\begin{equation}
 P_\mu^{\pm} = \frac{1}{2} \gamma_\mu 
  \quad (\mu = d), \qquad
 P_\mu^{\pm} = \frac{1}{2} ( \gamma_\mu \pm i r \gamma_d)
  \quad (\mbox{otherwise})
\end{equation}
 \item Bori\c{c}i--Creutz fermion \cite{Creutz:2007af,Borici:2007kz}
 \begin{equation}
  P_\mu^\pm = \frac{1}{2} 
   \left(
    (1 \pm ir) \gamma_\mu \mp ir \Gamma
   \right), \qquad
   \Gamma = \frac{1}{2} \sum_{\mu=1}^d \gamma_\mu
 \end{equation}
\end{itemize}
All these fermions include a free parameter $r$.
We often consider the case of $r=1$ for simplicity.
Note that another lattice fermion, which is called the staggered
fermion \cite{Kogut:1974ag,Susskind:1976jm,Sharatchandra:1981si}, is
essentially equivalent to the naive fermion, and also the $\pi$-flux
state as discussed in section~\ref{sec:tight}.
The naive and staggered fermions are transformed to each other by the
spin-diagonalization.
Thus in this paper we do not deal with the staggered fermion explicitly.

The Dirac equation, which corresponds to the Schr\"odinger
equation (\ref{Schrodinger}), is given by
\begin{equation}
 \sum_{\mu=1}^{d}
  \left[
   U_{\mu}(x) P^+_\mu \psi_{x+\hat\mu}
   - U_{\mu}^\dag(x-\hat\mu) P^-_\mu \psi_{x-\hat\mu}
  \right]
  = \lambda \psi_{x}.
  \label{Dirac_eq}
\end{equation}
When we write this equation (\ref{Dirac_eq}) as $D\psi = \lambda \psi$,
although this Dirac operator $D$ becomes non-Hermitian in general, one
can define an alternative Hermitian operator, $H=\gamma_{d+1} D$.
In the following we consider the spectrum of this Hermitian operator as
$H \psi = E \psi$, instead of the original Dirac equation (\ref{Dirac_eq}).

We can also obtain the corresponding Harper's equation by
taking Fourier transformation.
In this case, Harper's equation is slightly modified from
(\ref{Harper}) as
\begin{eqnarray}
 &&
  \sum_{s=1}^{d/2} 
  \gamma_{d+1}
  \Big[
     P^+_{2s} \tilde \psi_{z+\hat{s}} 
   - P^-_{2s} \tilde \psi_{z-\hat{s}}
   \nonumber \\
 &&
   + \Big(
    i \sin \left( k_{2s-1} - \omega_s z_s \right) 
    \left( P_{2s-1}^+ + P_{2s-1}^- \right)
    + \cos \left( k_{2s-1} - \omega_s z_s \right) 
    \left( P_{2s-1}^+ - P_{2s-1}^- \right)    
     \Big)
   \tilde \psi_z
  \Big]
  = E \tilde \psi_z .
  \nonumber \\
  \label{Harper_Dirac}
\end{eqnarray}
Here we have $2N_q$ energy bands due to the $\gamma$-matrix structure:
they appear in a pair for the relativistic theory.
In this case, it is again difficult to obtain a fully gapped spectrum
since this reduced equation is not one-dimensional, but
$d/2$-dimensional.

\section{Non-Abelian gauge field models}\label{sec:nonAbelian}

We consider another generalization of the Hofstadter problem in
higher dimensions by applying a non-Abelian gauge field as a background
configuration.
In this paper we concentrate on the four-dimensional model with $\SU(2)$
gauge field.
A further generalization to higher dimensional, and higher rank systems
seems to be straightforward.

\subsection{Background configuration}

We now introduce the following $\SU(2)$ background configuration as a simple
generalization of the $\U(1)$ background (\ref{gauge_pot}),
\begin{equation}
 A_0 = 0 , \qquad
 A_j = - \omega_j x_0 \sigma_j 
 \quad \mbox{for} \quad j = 1, 2, 3 ,
 \label{gauge_pot_NA}
\end{equation}
with $\sigma_j$ being the Pauli matrix.
Here we define filling fractions
\begin{equation}
 \omega_j = \frac{2\pi}{L^2} n_j \equiv 2\pi \frac{p_j}{q_j} .
\end{equation}
From this configuration we compute a field strength,
\begin{equation}
 F_{0j} = - \omega_j \sigma_j, \qquad
 F_{jk} = - \omega_j \omega_k x_0^2 \epsilon_{ijk} \sigma^i .
\end{equation}
The total background flux is given by
\begin{equation}
 \frac{1}{16\pi^2 }\int_{L^4} d^4 x \, \epsilon^{\mu\nu\rho\sigma} \, 
  \Tr F_{\mu\nu} F_{\rho\sigma} 
 = \frac{6\pi^2}{L^6} N \int_{L^4} d^4 x \, x_0^2
 = 2 \pi N .
\end{equation}
Here this integral is taken over the four-dimensional hypercubic lattice
of the size $L^4$, and we define the flux number $N = n_1 n_2 n_3$.

Thus the link variables associated with (\ref{gauge_pot_NA}) are given
as follows,
\begin{equation}
 U_0(x) = \id, \qquad
 U_j(x) = e^{-i\omega_j x_0 \sigma_j}
 = \id \cos \omega_j x_0 - i \sigma_j \sin \omega_j x_0 .
 \label{non_Abelian_conf}
\end{equation}
In this case, the translation symmetry of the lattice system yields
\begin{equation}
 x_0 \sim x_0 + q_{\rm LCM}, \qquad x_j \sim x_j + 1 ,
\end{equation}
where $q_{\rm LCM}$ is the least common multiple of $q_1, q_2$ and
$q_3$.
This means that the unit cell of this system is also extended in only
one dimension as well as the two-dimensional case with $\U(1)$ magnetic field.

\subsection{Lattice fermion models}

Let us first study the tight-binding model with the non-Abelian gauge
field configuration.
With a Fourier basis for $x_0 = q_{\rm LCM} X_0 + \bar{x}_0$, the
Hamiltonian (\ref{Ham}) gives rise to the corresponding Harper's
equation (\ref{Harper}),
\begin{equation}
 \tilde \psi_{\bar{x}_0+1} + \tilde \psi_{\bar{x}_0-1}
 + 2 \sum_{j=1}^3 
 \Big(
  \cos k_j \cos \omega_j \bar{x}_0 + \sigma_j \sin k_j \sin \omega_j \bar{x}_0
 \Big)
 \tilde \psi_{\bar{x}_0}
 = E \tilde \psi_{\bar{x}_0}.
 \label{Harper_SU2}
\end{equation}
This is just an effective one-dimensional model as the standard
two-dimensional Hofstadter problem.
The original four-dimensional model is reduced to this due to 
$\SU(2) \cong S^3$ gauge symmetry.
On the other hand, in this case it is also difficult to show its
spectrum is totally gapped as the case of the $\U(1)$ background in
higher dimensions.
Therefore we now study its density of states instead of the original
energy spectrum.

\begin{figure}[t]
 \begin{center}
  \includegraphics[width=30em]{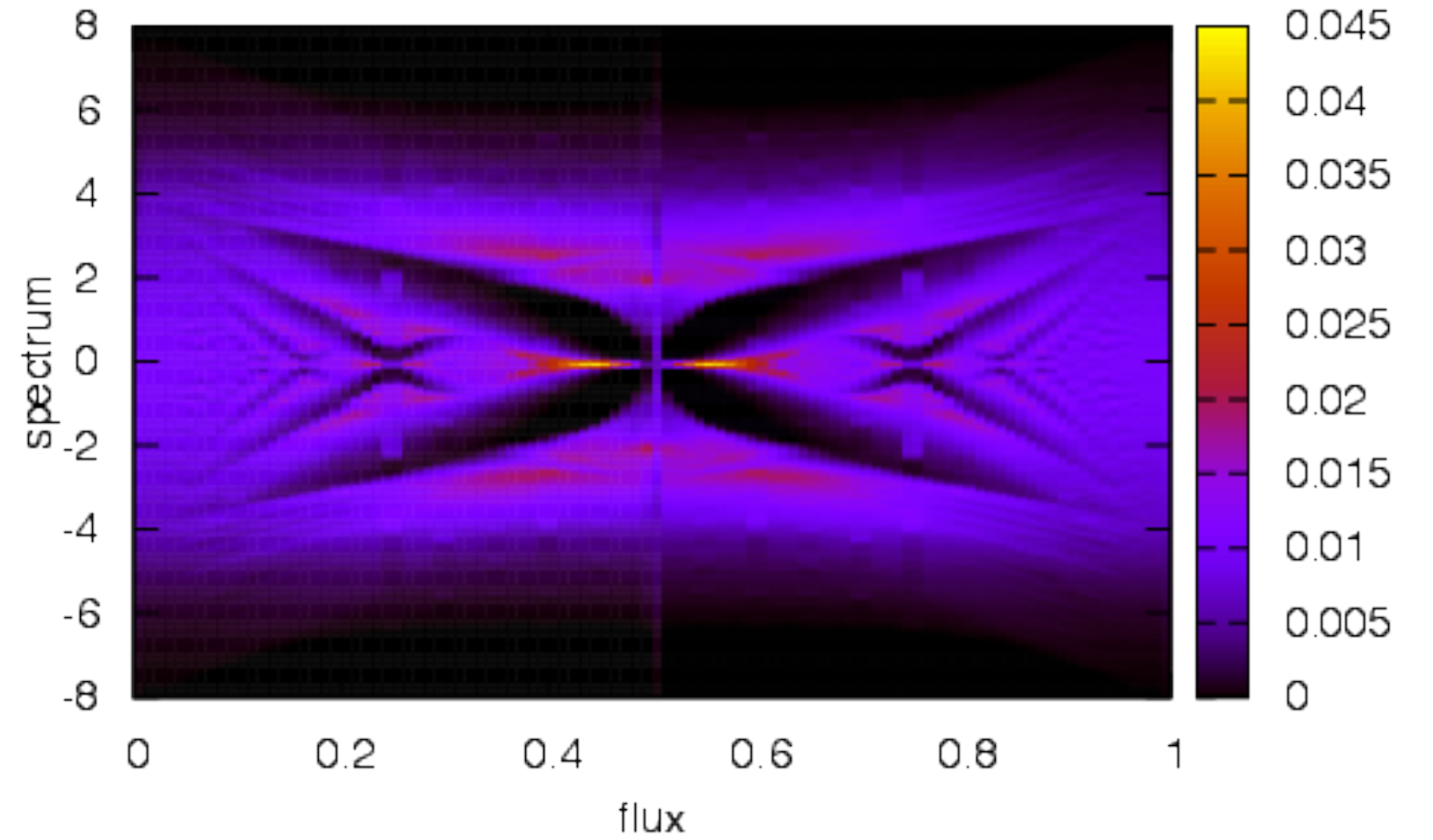}
 \end{center}
 \caption{Density of states for Harper's equation (\ref{Harper_SU2})
 of the flux 
 $\omega_{j=1, 2, 3} = \omega_0 \equiv 2\pi p/q$ with $q=100$, $p=0, 1, 2,
 \cdots 99$.}
 \label{spectrumDOS}
\end{figure}

Fig.~\ref{spectrumDOS} shows the density of states for Harper's
equation (\ref{Harper_SU2}) against $\omega_0$, with homogeneous
configuration $\omega_j=\omega_0$ for $j=1, 2, 3$.
Although its spectrum is not totally gapped, one can observe a
hierarchical structure of the spectrum, which seems fractal at least
based on this numerical computation.
In order to numerically determine the corresponding fractal dimension,
it is necessary to perform the calculation with a larger size system.
At the even fraction flux, e.g. $\omega_0/(2\pi) = 1/2, 1/4, \cdots$, we
find a dip in the spectrum.
Similar structure is discussed in the ordinary two-dimensional
Hofstadter problem at the gapless point of the filling fraction.

\begin{figure}[t]
 \begin{center}
  \includegraphics[width=19em]{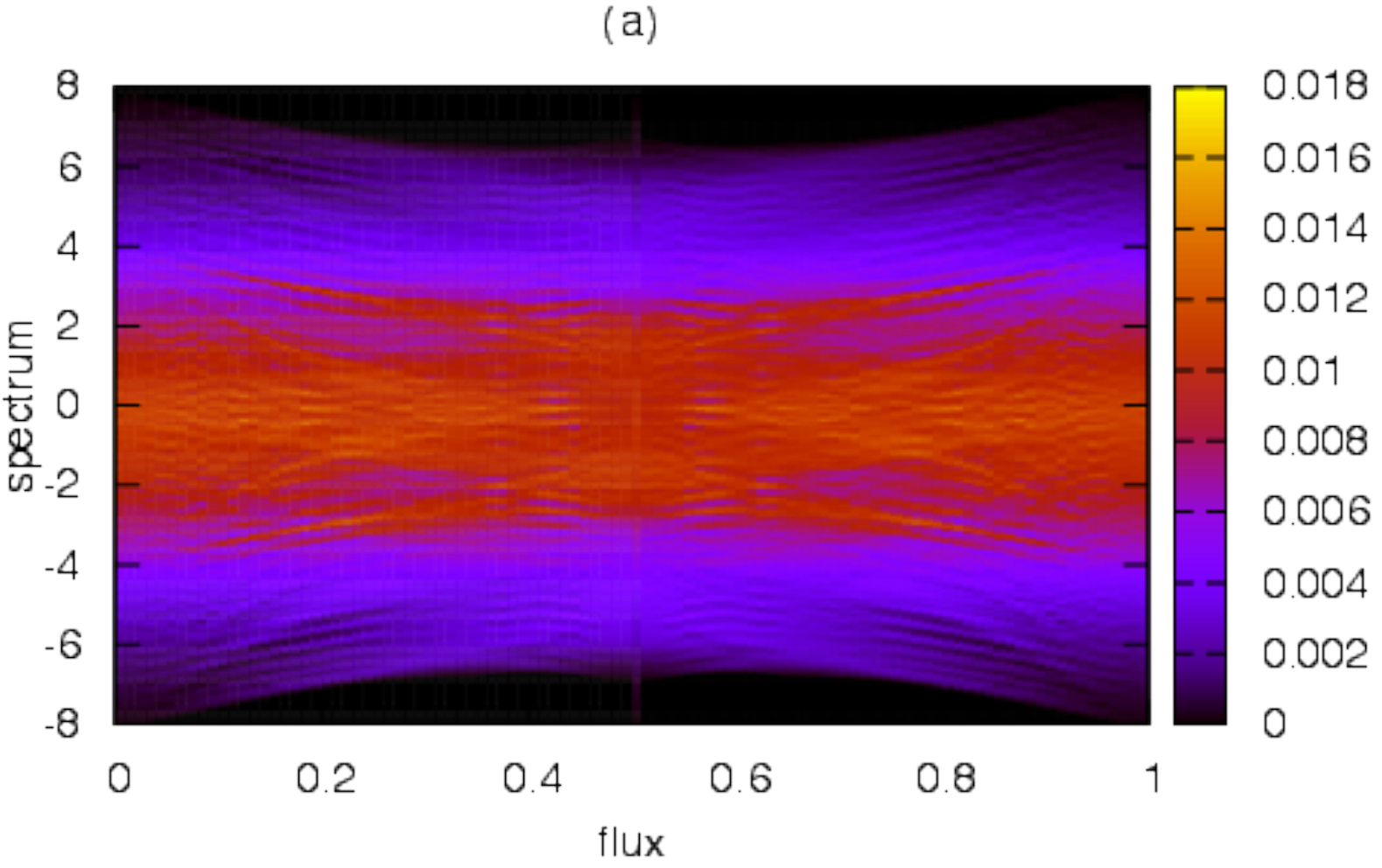}
  \includegraphics[width=19em]{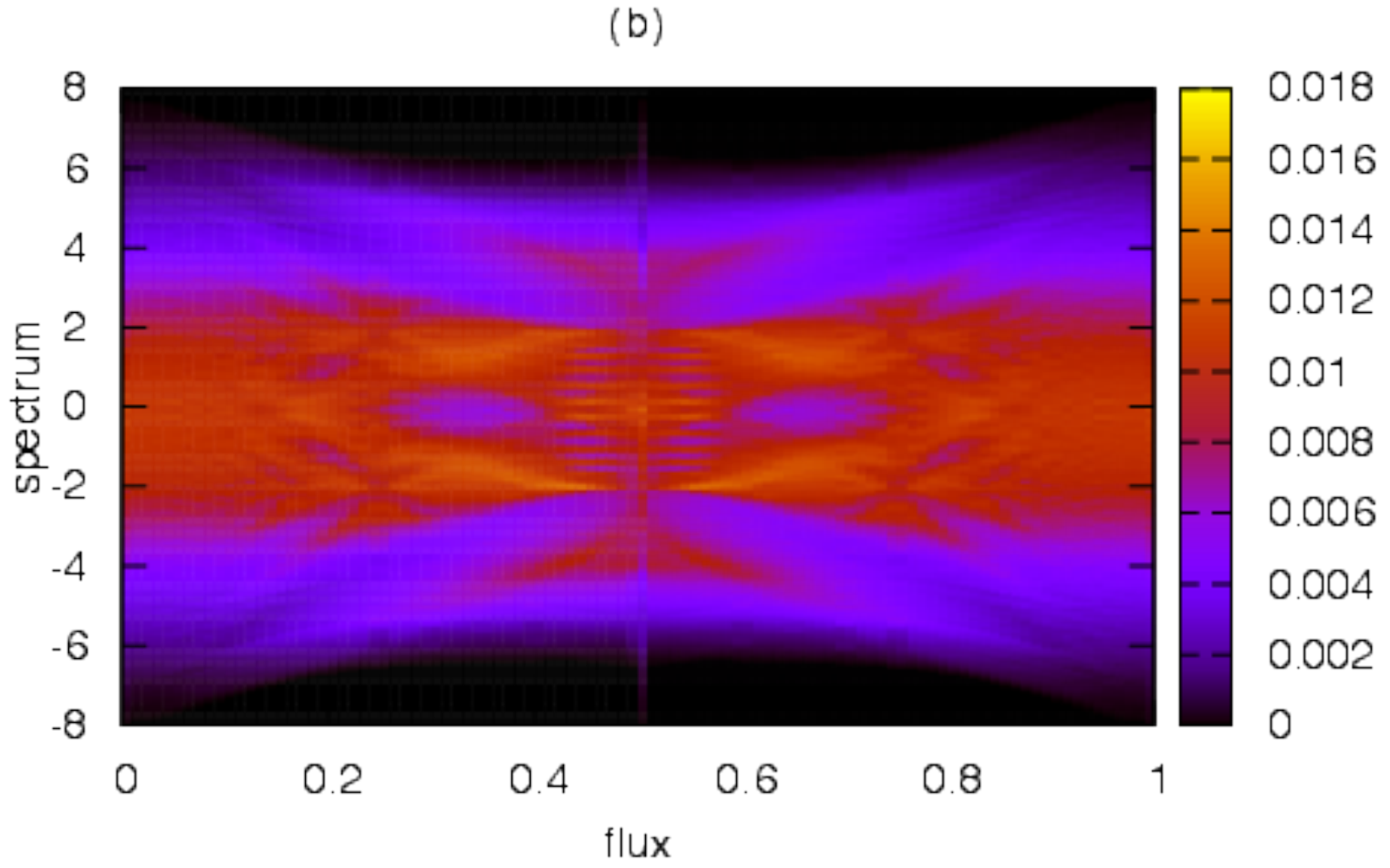}\\
  \vspace{1em}
  \includegraphics[width=19em]{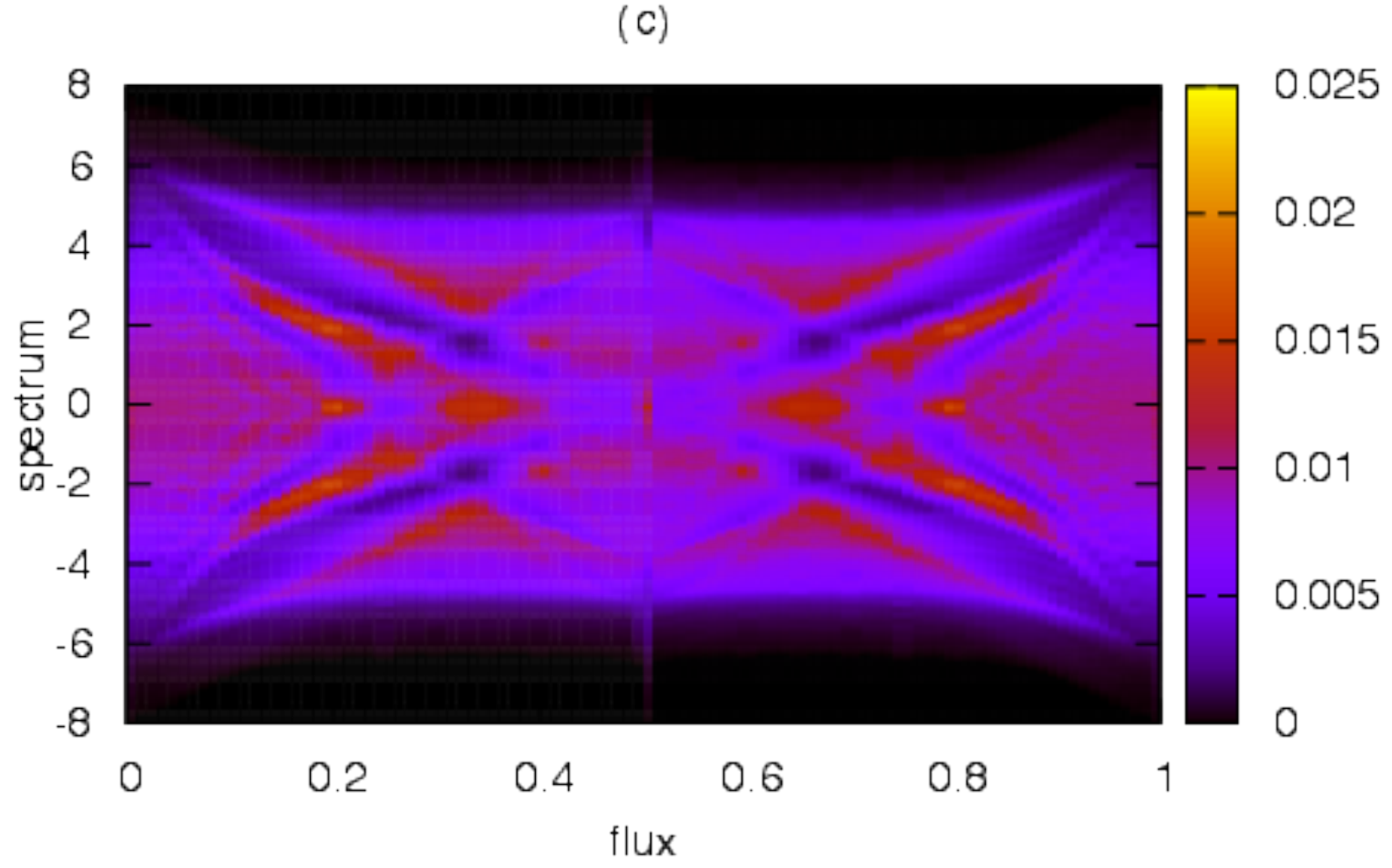}
  \includegraphics[width=19em]{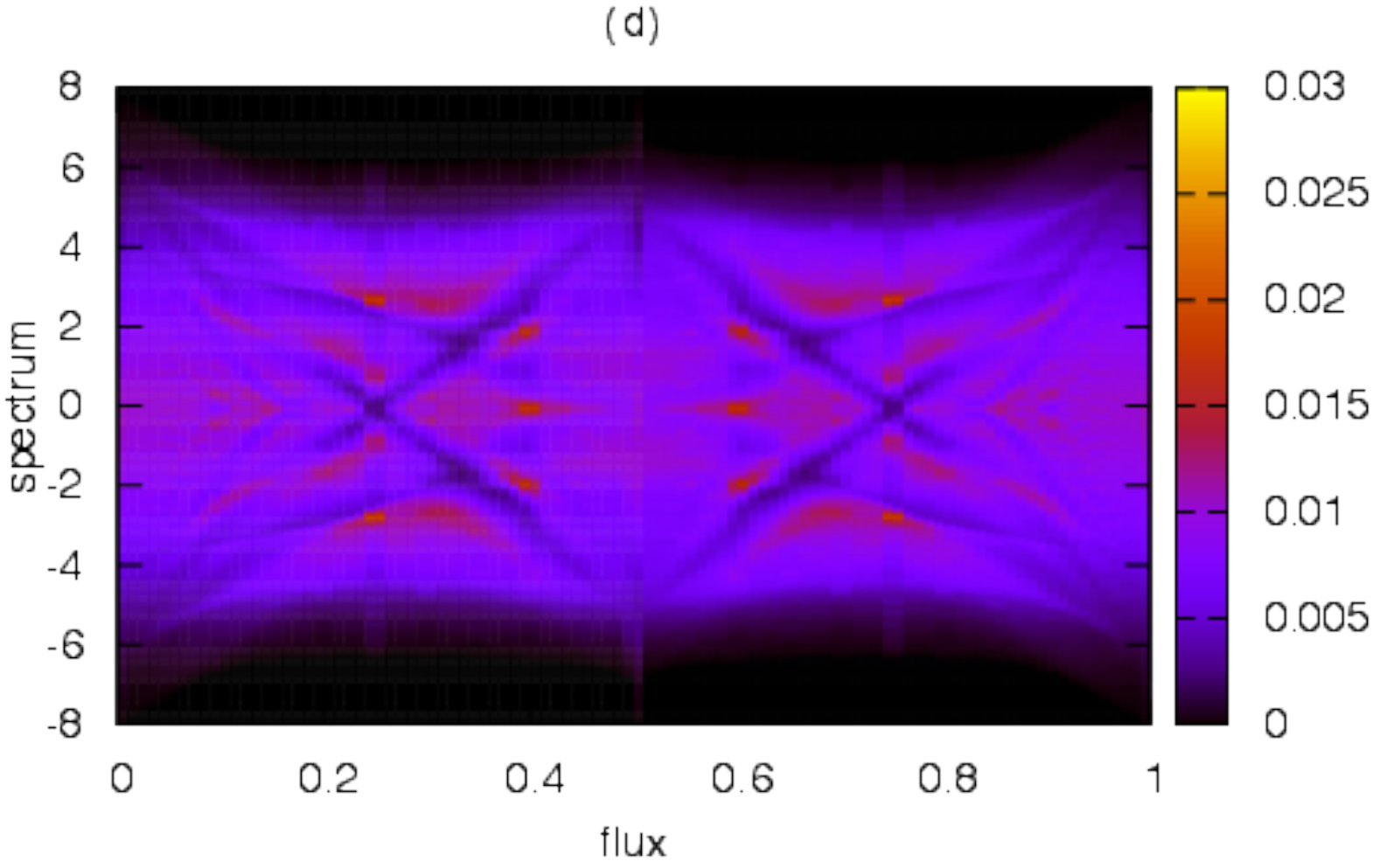}
 \end{center}
 \caption{Density of states for Harper's equation (\ref{Harper_SU2})
 with the inhomogeneous flux for $\vec{\omega} = \omega_0 \vec{n}$:
 (a) $\vec{n}=(1,0,0)$, (b) $(1,1,0)$, (c) $(2,1,1)$, and (d) $(2,2,1)$.}
 \label{spectrumDOS2}
\end{figure}

In Fig.~\ref{spectrumDOS2} we show the spectrum in the inhomogeneous
flux with fixing ratios of $\omega_{j=1,2,3}$.
When flux in some directions is turned off,
e.g. $\vec{\omega}=\omega_0 (1,0,0)$ and $\omega_0 (1,1,0)$, the
characteristic structure of Hofstadter's butterfly cannot be
observed in the spectrum: gap is completely closed.
In the cases of $\vec{\omega}=\omega_0 (2,1,1)$ and $\omega_0 (2,2,1)$,
there are some gaps in the spectrum and the hierarchical structure.
We also find a dip at the gapless point, $\omega_0=\pi$.

Let us now comment on the relation to the non-commutative space.
We introduce the following non-commutative torus,
\begin{equation}
 U V_j = e^{i\omega_j \sigma_j} V_j U
 \qquad \mbox{for} \quad j = 1, 2, 3.
 \label{NC_torus2}
\end{equation}
Here $U$ is defined as well as (\ref{NC_torus}), while $V_{j=1,2,3}$ are
slightly modified as 
\begin{equation}
 V_j = \mathrm{diag}\left(1, \, e^{i\omega_j \sigma_j}, \, \cdots, \,
		     e^{(q_{\rm LCM}-1)i\omega_j \sigma_j} \right).
\end{equation}
In this case the non-commutative parameter is matrix valued.
Thus, the commutation relations between $V_{j=1,2,3}$
cannot be written in a simple way.

As discussed in Sec.~\ref{sec:Abelian}, the tight-binding Hamiltonian
with the external field, corresponding to Harper's equation
(\ref{Harper_SU2}), can be associated with the non-commutative torus.
The naive form of the Hamiltonian without momentum dependence is given by
\begin{equation}
 U + U^\dag + \sum_{j=1}^3 
  \left[
   V_j + V_j^\dag
  \right].
\end{equation}
However, since the second part is represented as
\begin{equation}
 V_j + V_j^\dag = \mathrm{diag}
  \left(
   2 , \, 2 \cos \omega_j, \, \cdots, \, 2\cos (q_{\rm LCM}-1) \omega_j
  \right) \times \id ,
\end{equation}
we cannot involve a matrix structure in this way.
Actually the $\SU(2)$ nature of the external field is coming through the
momentum dependence in (\ref{Harper_SU2}).

We then investigate the Dirac fermion models as discussed in
section~\ref{sec:Dirac}.
Applying the background configuration (\ref{non_Abelian_conf}), the
corresponding Dirac equation is given by
\begin{eqnarray}
 && \gamma_5 
  \Big[
  P_0^+ \tilde \psi_{\bar{x}_0+1} - P_0^- \tilde \psi_{\bar{x}_0-1}
  \nonumber \\
 & &
  + \sum_{j=1}^3 
  \Big(
  \left(
   \cos k_j \cos \omega_j \bar x_0 + \sigma_j \sin k_j \sin \omega_j
   \bar x_0
  \right)
  \left( P_j^+ - P_j^- \right)
  \nonumber \\
 && \hspace{2.5em}
  +
  \left(
   \sin k_j \cos \omega_j \bar x_0 - \sigma_j \cos k_j \sin \omega_j
   \bar x_0
  \right)
  \left( P_j^+ + P_j^- \right)
  \Big) 
  \tilde \psi_{\bar x_0} \Big]
  = E \tilde \psi_{\bar x_0} .
\end{eqnarray}
Here we again show the Hermitian version of the Dirac equation by
multiplying the matrix $\gamma_5$.
When we write this Dirac operator in a matrix form, its
matrix size is $8q_{\rm LCM}=2$~(color) $\times$ 4~(spinor) $\times$
$q_{\rm LCM}$~(flux).
Here {\em color} corresponds to the rank of the gauge flux of $\SU(2)$.
On the other hand, the number of eigenvalues is given by $4q_{\rm LCM}$,
since each spectrum is doubly degenerated due to the spinor structure.



\section{Summary and discussions}\label{sec:summary}

In this paper we have explored some extensions of the Hofstadter problem
in higher dimensions.
First example is formulated with Abelian gauge configuration in $d=2r$, giving
rise to non-zero topological number.
We have shown that half of the gauge potential can be trivial by
applying the Landau gauge, thus the corresponding Harper's equation is
essentially written as $r$-dimensional lattice model.

We have also pointed out that the $\pi$-flux state is equivalent to the
staggered formalism of the relativistic lattice fermion.
The latter is directly related to the naive Dirac fermion through the
spin diagonalization.
This means the $\pi$-flux state involves the chiral symmetry, and thus
it yields massless excitation in any dimensions.

We have then investigated $\SU(2)$ non-Abelian gauge field configuration in
four dimensions.
We have considered the configuration with one specific direction in four
dimensions.
In this case hopping terms for the other three directions are reduced
due to $\SU(2)\cong S^3$ symmetry of the gauge field.
Thus we have obtained the one-dimensional Harper's equation by utilizing
the Fourier basis.
We have calculated its spectrum numerically, and its hierarchical
structure is actually observed.

Let us now comment on possibilities of future works along this
direction.
The two-dimensional Hofstadter problem is essentially
related to the quantum group
\cite{Wiegmann:1994zz,1994:WiegmannMPLB,PhysRevB.53.9697,PhysRevB.73.235118}:
Harper's equation is directly regarded as the Baxter's equation for
the one-dimensional $U_q(\mathfrak{sl}_2)$ model.
Thus it is interesting to explore the corresponding quantum group
structure to the generalized Hofstadter problems discussed in this paper.
In particular, the non-Abelian version of Harper's equation includes the
matrix-valued coefficient.
This corresponds to $q$-parameter in the two-dimensional case, thus it
is natural to investigate a kind of quantum group with matrix-valued
$q$-parameter.

Next is the lattice study with various kinds of lattice fermions, i.e. Wilson
\cite{Wilson:1974sk}, staggered
\cite{Kogut:1974ag,Susskind:1976jm,Sharatchandra:1981si},
staggered-Wilson
\cite{Adams:2009eb,Adams:2010gx,Hoelbling:2010jw,Creutz:2011cd,Misumi:2012sp},
minimal-doubling
\cite{Karsten:1981gd,Wilczek:1987kw,Creutz:2007af,Borici:2007kz,Kimura:2009qe,Kimura:2009di,Creutz:2010cz},
domain-wall \cite{Kaplan:1992bt,Shamir:1993zy,Furman:1994ky} and overlap
fermions \cite{Neuberger:1998wv}.
They were originally introduced to tackle the difficulty of the chiral
fermion on the lattice, but these formalisms themselves are interesting as
statistical lattice models: some of them are actually investigated in
the context of condensed-matter physics, for example, graphene, 
$\pi$-flux state, topological insulator/superconductor and so on.
Thus we hope the Hofstadter problem formulated with these lattice
fermions are relevant to realistic condensed-matter physics.

It is also definitely interesting to consider implications of the result
obtained in this paper for realistic situations.
An important difference between the non-Abelian gauge field and the
$\U(1)$ magnetic field is whether it breaks the time-reversal symmetry
of the system: the $\SU(2)$ gauge potential can be applied without
breaking the time-reversal symmetry.
It implies that, as a consequence of the dimensional reduction of the
model discussed in this paper, one can possibly realize the non-Abelian
Hofstadter system, for example, in topological insulators whose
time-reversal symmetry is not broken.
In addition, since there are already experimental techniques to realize the
non-Abelian gauge potential and the ordinary Hofstadter system based on
the $\U(1)$ field, respectively, one can expect that experimental
realization of the non-Abelian Hofstadter system might be possible
especially in the cold atomic system.

\subsection*{Acknowledgments}

The author would like to thank M.~Creutz, Y.~Hidaka and T.~Misumi for useful
discussions and comments.
The author is grateful to H.~Iida for collaboration at the early stage
of this work.
The author is supported by Grant-in-Aid for JSPS Fellows (No.~23-593).

\appendix

\section{Spin diagonalization}\label{sec:spin_diag}

We now show that there is an alternative expression of the
$d$-dimensional naive Dirac fermion without spinor matrix structure.
It is given by diagonalizing the corresponding $\gamma$ matrices.

Let us start with the naive Dirac fermion on the lattice,
\begin{equation}
 \mathcal{S} = \sum_{x,\mu}
  \left[
   \frac{1}{2} \bar\psi_x \gamma_\mu
   \left( \psi_{x+\hat\mu} - \psi_{x-\hat\mu} \right)
   + m \bar \psi_x \psi_x
  \right] .
  \label{naive_Dirac}
\end{equation}
Then, introducing the field $\chi_x$ defined as
\begin{equation}
 \psi_x = \gamma_d^{x_d} \cdots \gamma_2^{x_2} \gamma_1^{x_1}
  \chi_x, \qquad
 \bar \psi_x = \bar \chi_x 
 \gamma_1^{x_1} \gamma_2^{x_2} \cdots \gamma_d^{x_d} ,
\end{equation}
we can represent the naive lattice fermion (\ref{naive_Dirac}) in the
following form,
\begin{equation}
 \mathcal{S} = \sum_{x,\mu} 
  \left[
   \frac{1}{2} \eta_\mu(x) 
   \bar \chi_x \left( \chi_{x+\hat\mu} - \chi_{x-\hat\mu} \right)
   + m \bar \chi_x \chi_x
  \right] 
  \qquad \mbox{with} \qquad
  \eta_\mu(x) = (-1)^{\sum_{\nu<\mu} x_\nu} .
\end{equation}
Remark there is no spinor structure in this expression.
In other words, the spinor matrix is diagonalized in this basis.
This means the naive Dirac fermion can be rewritten in terms of
one-component fermionic field.
This is just the staggered formalism of the relativistic
fermion~\cite{Susskind:1976jm,Sharatchandra:1981si}.


\bibliographystyle{ytphys}
\bibliography{/Users/k_tar/Dropbox/etc/conf}

\end{document}